\documentclass[aps,draft,twocolumn]{revtex4}

\usepackage{bm}

\input epsf

\begin{document}

\title{
Electronic properties of the armchair graphene nanoribbon.
}
\author{A.V. Rozhkov$^{1, 2}$, S. Savel'ev$^{\rm 2,3}$,
Franco Nori$^{\rm 2,4}$}

\affiliation{
$^{1}$ Institute for Theoretical and Applied Electrodynamics Russian
Academy of Sciences, 125412 Moscow, Russia
}
\affiliation{
$^{3}$  Advanced Science Institute, The Institute of
Physical and Chemical Research (RIKEN), Wako-shi, Saitama, 351-0198, Japan 
}
\affiliation{
$^{3}$ Department of Physics, Loughborough University,
Loughborough LE11 3TU, UK 
}
\affiliation{
$^{4}$ Department of Physics, Center for Theoretical Physics,
Applied Physics Program, Center for the Study of Complex Systems,
The University of Michigan, Ann Arbor, MI 48109-1040, USA
}
%\maketitle

\begin{abstract}
We investigate the electronic band structure of an undoped graphene
armchair nanoribbon. We demonstrate that such nanoribbon always has a gap
in its electronic spectrum. Indeed, even in the situations where simple
single-electron calculations predict a metallic dispersion, the system is
unstable with respect to the deformation of the carbon-carbon bonds dangling
at the edges of the armchair nanoribbon. The edge bonds' deformation couples
electron and hole states with equal momentum. This coupling opens a gap at
the Fermi level. In a realistic sample, however, it is unlikely that this
instability could be observed in
its pure form. Namely, since chemical properties of the dangling carbon atoms
are different from chemical properties of the atoms inside the sample (for
example, the atoms at the edge have only two neighbours, besides additional
non-carbon atoms might be attached to passivate unpaired covalent carbon
bonds), it is very probable that the bonds at the edge are deformed due to
chemical interactions. This chemically-induced modification of the
nanoribbon's edges can be viewed as an effective field biasing our
predicted instability in a particular direction. Yet by disordering this
field (e.g., through random substitution of the radicals attached to the
edges) we may tune the system back to the critical regime and vary the
electronic properties of the system. For example, we show that electrical
transport through a nanoribbon is strongly affected by such disorder.
\end{abstract}

\date{\today}

\maketitle
\hfill
%\draft

\section{Introduction}

Graphene is attracting considerable attention due to its unusual electronic
properties including: large mean free path, ``relativistic" dispersion of the
low-lying electron states, and ``valley" degeneracy 
\cite{neto_etal}.
These remarkable features made many researchers hope that some day graphene
mesoscopic structures might revolutionise nanoscience. Thus, a substantial
amount of effort has been invested investigating graphene devices,
such as quantum dots
\cite{dot_reference}, 
bilayer structures
\cite{2-layer},
and nanoribbons
\cite{nanoribbon_reference}.

Studying the physics of nanoribbons, a certain discrepancy between results of
first-principle calculations
\cite{louieI,louieII},
experiments 
\cite{gap_experimentI,gap_experimentII},
and single-electron approximations
\cite{single_eI,single_eII,single_eIII,single_eIV} was stumbled upon: 
whereas the single-electron approximation predicts that, depending on its
width, an armchair nanoribbon could be either semiconducting or metallic, both
experiments and first-principle calculations suggest that it is always
semiconducting: a nanoribbon of any width $W$ has gap at
the Fermi level; the magnitude of the gap scales as $1/W$.

Two different mechanisms were proposed to explain this disagreement.
According to the first mechanism the gap is due to electron-electron
interactions \cite{sandler}. At zero doping the interaction induces a charge
gap. The physics here is similar to the Mott transition in the Hubbard model
at half filling. If the Coulomb coupling constant $g$ is large 
($g \gtrsim t$, 
where $t$ is the carbon-carbon hopping amplitude), the gap scales as $g$. If
$g$ is small, the gap vanishes faster than $g$. Since 
$g \propto 1/W$,
such mechanism is consistent with the observed scaling 
$\Delta \propto 1/W$ 
for small $W$ only. For large $W$ the gap decays faster than $1/W$.

\begin{figure}
\centering
\leavevmode
\epsfxsize=8.5cm
\epsfysize=5.8cm
\epsfbox{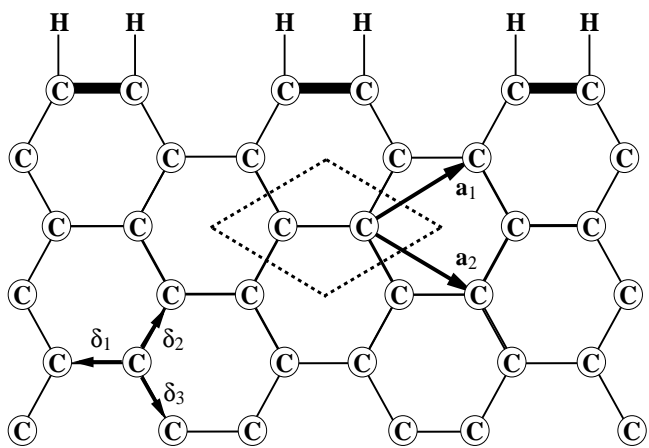}
\caption[]
{\label{edge-H}
A schematic diagram of a graphene sheet, with one edge passivated by hydrogen.
Deformed bonds are shown by bold lines, atoms of carbon and hydrogen are
represented by the symbols `C' and `H'. The vectors 
${\bm \delta}_{1,2,3}$ 
connect nearest neighbours on the graphene lattice. The vectors
${\bm a}_{1,2}$ 
are the primitive lattice vectors. The broken line diamond is the graphene
unit cell.
}
\end{figure}

Another way to explain the gap was outlined in
\cite{louieI},
where first-principle computations have shown that the length of the
carbon-carbon bonds at the hydrogen-passivated edge is shorter than the
length of the bonds in the bulk. Due to this, the hopping amplitude across
the bonds dangling from the nanoribbon edge
($t_{\rm edge}$) 
differs from $t$:
$t_{\rm edge} = t + \delta t$, 
$\delta t \sim 0.1 t$
(see Fig.\ref{edge-H} where the short bonds are shown in bold). 
Because of
$\delta t$,
the nanoribbon's Hamiltonian acquires an additional term. It couples states
with the same momentum above and below the Fermi energy; such coupling
opens a gap in the electronic spectrum. Assuming that
$\delta t$ 
depends mainly on the chemical properties of the edge (e.g., nature of the
passivating radical or
absence thereof), one can demonstrate that the effective strength of this
coupling is inversely proportional to $W$. Thus, the relation
$\Delta \propto 1/W$ 
is recovered.

In addition to these mechanisms, other instabilities can be present as well.
Which mechanism of gap generation dominates depends on a variety of factors,
such as electrostatic screening by the gate electrode, mechanical forces
applied to the nanoribbon, edge passivation, etc.

\subsection{Summary of our results}

In this paper we will study in detail the second mechanism described above.
We will show that the armchair nanoribbon is unstable toward the deformations
of the edge bonds. Specifically, we prove the following: if the radical
passivating the edge is carefully picked (to ensure that it does not modify
the hopping at the edges: 
$t_{\rm edge} = t$),
then the total energy of the nanoribbon would become a decreasing function
of the edge bond deformation. In other words, spontaneous deformations of the
dangling bonds are favourable for they decrease the total nanoribbon energy.
Such deformations make
$t_{\rm edge}$
unequal to
$t$.

To understand the physics behind this, it is convenient to describe such 
deformation using a real one-component (Ising-like) order parameter
$\delta t$. 
At the mean-field level we can state that the energy of the deformed bonds is
increased by the amount
$\epsilon_{\rm b} \sim (\delta t)^2 > 0$,
while the energy of the conducting electrons decreases by the amount
$\epsilon_{\rm el} \sim \Delta^2 \ln (\Delta/t) < 0$. 
Since 
$\Delta \sim | \delta t |$,
the total energy,
$\epsilon_{\rm b} + \epsilon_{\rm el}$,
always has a minimum at not-zero 
$| \delta t |$. 

It is useful to extend the analogy between our system and the Ising model
further. Notice that our chemical forces deforming the edge bonds
are similar to the ``magnetic field" coupled to the order parameter (the
``magnetisation") in the Ising model. Thus, it is natural to refer to these
chemical forces as the ``edge field".

In practice, we have very limited control over the magnitude of the ``edge
field": it is very difficult to select the radical attached to the
nanoribbon's edge to guarantee that 
$\delta t = 0$.
Using the analogy between our nanoribbon and the Ising model, we conclude
that, since the ``field" is always switched on, our system is always away from
the critical regime.

Yet even a system with a strong ``edge field" may be tuned to criticality
by disordering the ``field". This way, the disorder may be used as a
tool to control spectral properties of nanoribbons.

Disorder can be introduced by passivating the edges by radicals of two
different types, randomly distributed along the length of the nanoribbon. The
hopping amplitude at the edge 
$t_{\rm edge}$
would become a function of the coordinate along the ribbon:
\begin{eqnarray}
t_{\rm edge} (x) = t 
		   +
		   {\overline{\delta t}} 
		   + 
		   \delta t_{\rm dis} (x),
%%%%%%%%%%%%%%%%%%%%%%%%%%%%%%%%%
\label{t_edge}%%%%%%%%%%%%%%%%%%%
%%%%%%%%%%%%%%%%%%%%%%%%%%%%%%%%%
\end{eqnarray}
where 
${\overline{\delta t}}$
is independent of $x$, while
$\delta t_{\rm dis} (x)$
is the disordered part.

In such system the electronic spectrum is very sensitive to the relative
strengths of
${\overline{\delta t}}$
and
$\delta t_{\rm dis}$:
when the disorder is weak, the spectrum has (pseudo)gap due to non-zero
${\overline{\delta t}}$
term; when 
${\overline{\delta t}}=0$,
the pseudogap closes, and the nanoribbon is in the disorder-dominated regime. 

The crossover between these two regimes can be observed with the help of 
transport measurements. In a nanoribbon whose length $L$ is thermodynamically
large, the conductance vanishes exponentially in both regimes: when there is
no disorder, the finite gap removes density of states from the Fermi level;
when there is disorder, the wave functions are localized. Thus, for very long
nanoribbons the conductance is suppressed regardless of the disorder strength.

The situation is different for a mesoscopic sample. Note that the
localization is a weaker phenomenon than the gap opening in the sense that
the inverse of the localization length 
$l_{\rm loc}^{-1}$ 
is the second order in the disorder strength:
$l_{\rm loc}^{-1}=O(\delta t^2_{\rm dis})$, 
while the inverse of the length scale
$l_{\rm gap}^{-1} \sim \Delta/ v_{\rm F}$
charachterizing the gap size is linear in 
$\delta t$:
$l_{\rm gap}^{-1} = O( |\delta t|)$. 
That is, a longer nanoribbon is required to observe a well-developed
localization. Therefore, it is possible to choose $L$ sufficiently short to
have no localization features at any disorder, and yet long enough to observe
the spectral gap when the disorder is low. Thus, when the disorder is absent,
the low-temperature conductance of the sample is exponentially suppressed due
to the gap; otherwise, the conductance is finite down to the lowest
temperatures. Therefore, the disorder effectively closes the gap.

The paper is organized as follows. In Sect. \ref{model} we derive the model
for a nanoribbon with deformed edges. The instability of the nanoribbon
toward the deformation of the edge bonds is discussed in Sect. 
\ref{instability}. In Sect. \ref{disorder} we investigate the effect of
disorder on the transport properties of the nanoribbon. Sect.
\ref{conclusions} presents the conclusions.

\section{Model}
\label{model}

In this section we will obtain the Hamiltonian for a graphene nanoribbon
with deformed edge bonds. Our derivation relies on basic facts of the graphene
physics which are discussed in Ref. \cite{neto_etal}.

\subsection{Tight-binding model for graphene}

For completeness, in this subsection we quickly rederive basic
single-electron properties of a graphene sheet. This gives us an opportunity
to introduce notation we will need below.

It is common to describe a graphene sample in terms of a tight-binding model
on a honeycomb lattice. Such lattice can be split into two sublattices,
denoted by
${\cal A}$
and
${\cal B}$.

The Hamiltonian of a graphene sheet is given by:
\begin{eqnarray}
H = 
-t
\sum_{\bf R} 
\sum_{i=1,2,3}
c^\dagger_{\bf R}
c^{\vphantom{\dagger}}_{{\bf R} + {\bm \delta}_i} 
+
{\rm H.c.},
%%%%%%%%%%%%%%%%%%%%%%%%%%%%%
\label{H}%%%%%%%%%%%%%%%%%%%%
%%%%%%%%%%%%%%%%%%%%%%%%%%%%%
\end{eqnarray}
where ${\bf R}$ runs over sublattice ${\cal A}$. The vectors 
${\bm \delta}_i$ ($i=1,2,3$) connect the nearest neighbours. These are
(see Fig. \ref{ribbon}):
\begin{eqnarray}
{\bm \delta}_1 
&=&
a_0 (-1, 0),
\\
{\bm \delta}_2 
&=& 
a_0 (1/2, \sqrt{3}/2)
\\
{\bm \delta}_3 
&=& 
a_0 (1/2, -\sqrt{3}/2).
\end{eqnarray}
The symbol $a_0$ denotes the carbon-carbon bond length, which is about 1.4
\AA.

The corresponding Schr\"odinger equation can be written as:
\begin{eqnarray}
\varepsilon \psi^{\cal A}_{\bf R} 
&=&
- t
\psi^{\cal B}_{{\bf R} + {\bm \delta}_1}
-
t \sum_{i=1,2}
	\psi^{\cal B}_{{\bf R} + {\bm \delta}_1 + {\bf a}_i},
%%%%%%%%%%%%%%%%%%%%%%%%%%%%%
\label{sch_a}%%%%%%%%%%%%%%%%
%%%%%%%%%%%%%%%%%%%%%%%%%%%%%
\\
\varepsilon \psi^{\cal B}_{{\bf R} + {\bm \delta}_1} 
&=& 
- t \psi^{\cal A}_{\bf R} 
- 
t \sum_{i=1,2}
	\psi^{\cal A}_{{\bf R} - {\bf a}_i},
%%%%%%%%%%%%%%%%%%%%%%%%%%%%%%%
\label{sch_b}%%%%%%%%%%%%%%%%%%
%%%%%%%%%%%%%%%%%%%%%%%%%%%%%%%
\end{eqnarray}
where 
$\psi_{\bf R}^{\cal A}$ 
($\psi_{{\bf R} + {\bm \delta}_1}^{\cal B}$) 
denotes the wave function value at the site 
${\bf R}$ 
(at the site ${\bf R} + {\bm \delta}_1$)
of sublattice ${\cal A}$ (sublattice ${\cal B}$).
The primitive vectors of the honeycomb lattice are:
\begin{eqnarray}
{\bf a}_1 &=& a_0 (3/2, \sqrt{3}/2),
\\
{\bf a}_2 &=& a_0 (3/2, -\sqrt{3}/2).
\end{eqnarray}
They connect nearest neighbours on the same sublattice. 

Since the primitive cell contains two atoms, it is convenient to define a
two-component (spinor) wave function:
\begin{eqnarray}
\Psi_{\bf R} 
= 
\left(
	\matrix{ 
			\psi_{\bf R}^{{\cal A} \hphantom{\delta_1}}\cr
			\psi_{{\bf R} + {\bm \delta}_1}^{\cal B}\cr
		}
\right).
\end{eqnarray} 
With this notation the action of $H$ on spinor
$\Psi_{\bf k}$
can be expressed as:
\begin{eqnarray}
H \Psi_{\bf k} 
=
\left(
	\matrix{ 
			0            &  -t_{\bf k}  \cr
       		       -t_{\bf k}^*  &        0     \cr
       		}
\right)
\Psi_{\bf k},
\\
t_{\bf k} 
=
t
\left[
	1 
	+ 
	2 {\exp}	\left(
				-i\frac{3  k_x a_0}{2} 
			\right)
	\cos \left(
			\frac{\sqrt{3}}{2} k_y a_0
	      \right)
\right].
\end{eqnarray}
For every ${\bf k}$ there are two eigenstates:
\begin{eqnarray}
\Psi_{{\bf k} \pm}
=
\left(
	\matrix{
			\mp {\rm e}^{i \theta_{\bf k}} \cr
				1			\cr
		}
\right),
\\
\exp 	\left(
	{i \theta_{\bf k}} 
	\right)
=
\frac{t_{\bf k}}{|t_{\bf k}|},
\end{eqnarray} 
with the eigenvalues:
\begin{widetext}
\begin{eqnarray}
\varepsilon_{{\bf k} \pm}
%&=&
=
\pm |t_{\bf k}|
%%%%%%%%%%%%%%%%%%%%%%%%%%%%%%%%%%
\label{graphene_energy}%%%%%%%%%%%
%%%%%%%%%%%%%%%%%%%%%%%%%%%%%%%%%%
%\\
%\nonumber
%&=&
=
\pm t
\sqrt{
	1 
	+ 
	4 \cos \left( 
			\frac{3}{2} k_x a_0 
		\right)
	  \cos \left( 
			\frac{\sqrt{3}}{2} k_y a_0 
		\right)
%\right.
%\\
%\nonumber 
%&&\left.
	+
	4 \cos^2 \left( 
			\frac{\sqrt{3}}{2} k_y a_0 
		\right)
}.
\end{eqnarray}
\end{widetext}
The states with negative (positive) energy are filled (empty) at $T=0$.

\begin{figure} 
\centering
\leavevmode
\epsfxsize=8.5cm
\epsfysize=8.8cm
\epsfbox{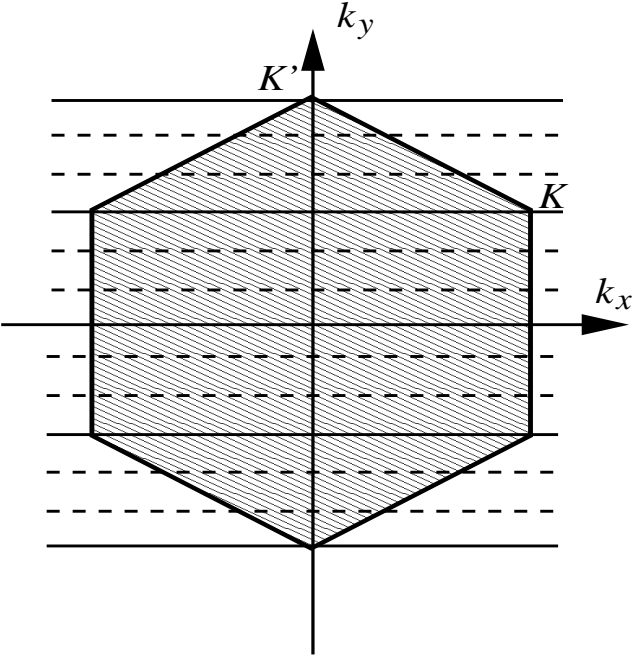}
\caption[]
{\label{bz}
Shaded hexagon schematically shows Brillouin zone of graphene. The Dirac
cones are located at the corners of the zone. In an armchair nanoribbon the
condition Eq.~(\ref{quantization}) imposes the quantization of $k_y$. This
splits the whole spectrum into a finite number of 1D branches. A
branch is represented by a horizontal line. Gapless branches (solid lines)
pass through Dirac cones whereas branches with the gap (broken lines) do not.
A given branch may be represented by more than one line on this figure [see
discussion before Eq.~(\ref{band_index})].
}
\end{figure}

The quantity 
$\varepsilon_{\bf k}$
vanishes at six points within the Brillouin zone (see Fig.~\ref{bz}):
$(0, \pm 4\pi/(3\sqrt{3}a_0))$ 
and
$(\pm 2\pi / (3 a_0), \pm 2\pi/(3\sqrt{3}a_0))$.
These are the locations of the famous Dirac cones of graphene.

These six cones can be split into two equivalence classes: the locations of
any two cones inside the same equivalence class differ by a reciprocal
lattice vector. Thus, we do not need all six of them. Two inequivalent cones
are sufficient:
\begin{eqnarray}
&\text{cone}& {\cal K}\hphantom{'}: {\bf k}_{{\cal K}\hphantom{'}} =
( 2 \pi / ( 3 a_0 ),  2 \pi / (3\sqrt{3} a_0));
%%%%%%%%%%%%%%%%%%%%%%%%%%%%%%%
\label{K}%%%%%%%%%%%%%%%%%%%%%%
%%%%%%%%%%%%%%%%%%%%%%%%%%%%%%%
\\
&\text{cone}& {\cal K}': {\bf k}_{{\cal K}'} =
(0, 4 \pi / (3\sqrt{3} a_0)).
%%%%%%%%%%%%%%%%%%%%%%%%%%%%%%%
\label{K'}%%%%%%%%%%%%%%%%%%%%%
%%%%%%%%%%%%%%%%%%%%%%%%%%%%%%%
\end{eqnarray}

\subsection{The nanoribbon spectrum}

\begin{figure} 
\centering
\leavevmode
\epsfxsize=8.5cm
\epsfysize=6.5cm
\epsfbox{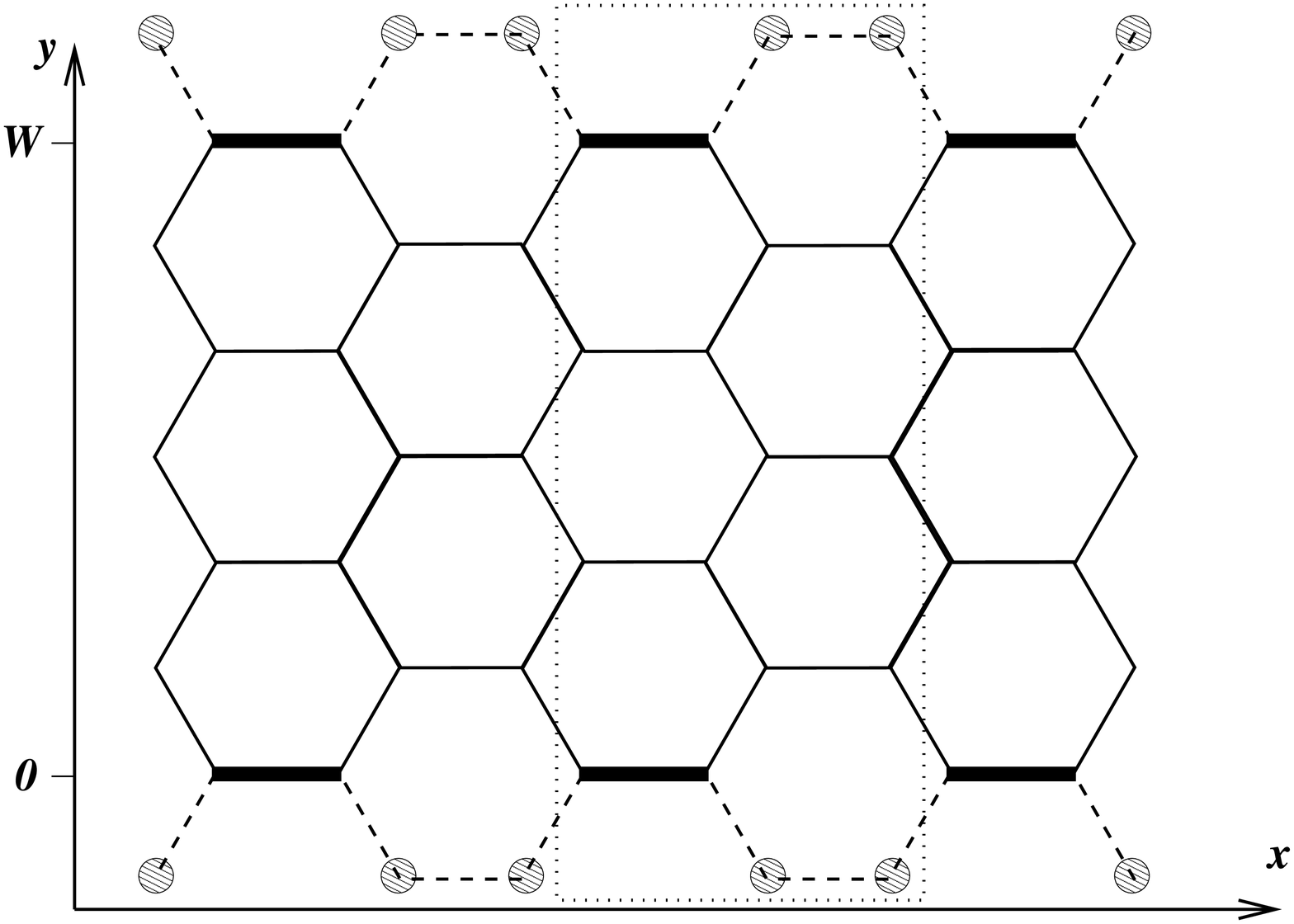}
\caption[]
{\label{ribbon}
A segment of an armchair nanoribbon of width $W$. The auxiliary sites at the
edges (where the wave function must vanish) are shown by the hatched circles.
The nanoribbon unit
cell is enclosed inside the dotted line. The deformed bonds at the edges
are shown bold.
}
\end{figure}

Consider now the electron states of the armchair nanoribbon (see Fig.~
\ref{ribbon}). 
Such nanoribbon is defined by the condition 
$0 \leq y \leq W$. 
The width of the nanoribbon $W$ is a multiple of 
$\sqrt{3} a_0 /2$:
\begin{eqnarray} 
W = \frac{\sqrt{3} a_0}{ 2}M,
\end{eqnarray} 
where $M$ is an integer.

The Schr\"odinger equation for the edge sites differs from Eqs.~(\ref{sch_a})
and (\ref{sch_b}). For sites on the upper edge 
($y=W$)
we can write:
\begin{eqnarray}
\varepsilon \psi^{\cal A}_{\bf R} 
&=&
- t
\psi^{\cal B}_{{\bf R} + {\bm \delta}_1}
-
t \psi^{\cal B}_{{\bf R} + {\bm \delta}_1 + {\bf a}_2},
%%%%%%%%%%%%%%%%%%%%%%%%%%%%%
\label{sch_a_edge}%%%%%%%%%%%
%%%%%%%%%%%%%%%%%%%%%%%%%%%%%
\\
\varepsilon \psi^{\cal B}_{{\bf R} + {\bm \delta}_1} 
&=& 
- t \psi^{\cal A}_{\bf R} 
- 
t \psi^{\cal A}_{{\bf R} - {\bf a}_1}.
%%%%%%%%%%%%%%%%%%%%%%%%%%%%%%%
\label{sch_b_edge}%%%%%%%%%%%%%
%%%%%%%%%%%%%%%%%%%%%%%%%%%%%%%
\end{eqnarray}
Note the absence of the summation over the lattice vectors in the
right-hand side of these equations. This boundary condition is called free.
It can be easily generalized for the lower edge.

An armchair nanoribbon is invariant under a shift over $3 a_0$ along the
$x$-axis. Thus, a $3 a_0$ long nanoribbon segment can be thought of as a
nanoribbon unit cell. There are $(M+1)$ graphene unit cells in a nanoribbon
unit cell. The nanoribbon Brillouin zone is:
\begin{eqnarray}
-\, \frac{\pi}{3 a_0} < k_x < \frac{\pi}{3 a_0}.
%%%%%%%%%%%%%%%%%%%%%%%%%%%%%%%%
\label{nanoribbon_bz}%%%%%%%%%%%
%%%%%%%%%%%%%%%%%%%%%%%%%%%%%%%%
\end{eqnarray}

Now we assume that all carbon-carbon bonds of our system have the same
hopping amplitude $t$ (that is, the deformation of the edge bonds is absent). 
The wave function $\Psi_{\bf R}$ satisfies the free boundary conditions
Eqs.~(\ref{sch_a_edge}) and (\ref{sch_b_edge}) at 
${\bf R} = (x, W)$
and similar conditions at
${\bf R} = (x, 0)$. 
It is not convenient, however, to work with such boundary conditions directly.
Fortunately, if we add an additional row of lattice sites at each edge and
demand that the wave function vanishes at these auxiliary sites (see Fig.~
\ref{ribbon}), then the wave function at the real edge sites satisfies the
free boundary condition. In other words, our {\it free} boundary condition
problem for a nanoribbon of width $W$ is equivalent to the {\it zero}
boundary condition problem for a nanoribbon of width 
$W + \sqrt{3} a_0$.
Thus, we want:
\begin{eqnarray}
\Psi_{\bf R} |_{y = - \sqrt{3} a_0 / 2 }
=
\Psi_{\bf R} |_{y =  W + \sqrt{3} a_0 / 2} = 0.
%%%%%%%%%%%%%%%%%%%%%%%%%%%%%
\label{bc}%%%%%%%%%%%%%%%%%%%
%%%%%%%%%%%%%%%%%%%%%%%%%%%%%
\end{eqnarray} 
The eigenfunction of the Hamiltonian $H$, Eq.~(\ref{H}), satisfying Eq.~
(\ref{bc}) can be written as follows:
\begin{eqnarray}
\Psi_{{\bf R} \pm} 
= 
\left(
	c_1
	\Psi_{k_x, k_y \pm} {\rm e}^{ - i k_y y}
	+
	c_2
	\Psi_{k_x, -k_y \pm} {\rm e}^{  i k_y y}
\right) 
{\rm e}^{ - i k_x x},
%%%%%%%%%%%%%%%%%%%%%%%%%%%%%
\label{nanoribbon_wf}%%%%%%%%
%%%%%%%%%%%%%%%%%%%%%%%%%%%%%
\end{eqnarray}
where $c_{1,2}$ are complex coefficients. Note that this eigenfunction has a
well-defined value of the momentum $k_x$ along the $x$-axis, but not of the
momentum $k_y$ along the $y$-axis, since our system has no translational
invariance in the $y$-direction. 

The values of $k_y$ and $c_{1,2}$ in Eq.~(\ref{nanoribbon_wf}) must
be chosen to satisfy Eq.~(\ref{bc}). Since the spinor 
$\Psi_{{\bf k}\pm}$ 
remains the same when the sign of $k_y$ changes, we derive: 
\begin{eqnarray}
\sin 
	\left[
		k_y (W + \sqrt{3} a_0)
	\right]
=
0,
\\
c_1 = - {\exp}	\left(
			{-i \sqrt{3} k_y a_0}
		\right)
	 c_2.
\end{eqnarray}
We then obtain the following quantization condition:
\begin{eqnarray}
k_y 
&=&
\frac{
	2 \pi n
     }
     {
	\sqrt{3}( M + 2 ) a_0
     },
%%%%%%%%%%%%%%%%%%%%%%%%%%%%%%%
\label{quantization}%%%%%%%%%%%
%%%%%%%%%%%%%%%%%%%%%%%%%%%%%%%
\end{eqnarray}
where $n$ is an integer. Thus, the nanoribbon spectrum 
$\varepsilon_n (k_x)$
consists of a set of one-dimensional branches labelled by an integer, $n$.

Using the above results it is possible to construct explicitly the
nanoribbon eigenfunction. However, for our purposes it is more convenient to
define an effective Hamiltonian for a given branch. Let
us look for a nanoribbon Hamiltonian eigenfunction in the form:
\begin{eqnarray}
\Psi_{\bf R} = \Psi_n (x) 
\sin 
	\left[
		\frac{ 2 \pi n }{ \sqrt{3} (M+2) a_0 }
		\left(
			y + \frac{\sqrt{3} a_0}{2}
		\right)
	\right],
\end{eqnarray}
where the spinor $\Psi (x)$ is defined as:
\begin{eqnarray}
\Psi_n (x)
=
\left(
	\matrix{	\psi_n^{\cal A} (x) \cr
			\psi_n^{\cal B} (x - a_0) 
		}
\right).
\end{eqnarray} 
In this equation
$x = 3 a_0 m/2$,
and $m$ is an integer.

Substituting the expression for 
$\Psi_n$
into the Schr\"odinger equations (\ref{sch_a}) and (\ref{sch_b}), we obtain:
\begin{widetext}
\begin{eqnarray}
\varepsilon \Psi_n (x)
= 
\left(
	\matrix{	0 & -t \cr
			-t&  0 
		}
\right)
\Psi_n (x)
+
2 \cos ( \sqrt{3} k_y a_0 / 2 )
\left[
	\left(
		\matrix{	0 & -t \cr
				0 &  0 
			}
	\right)
	\Psi_n (x + 3 a_0 /2 )
	+
	\left(
		\matrix{	0 & 0 \cr
				-t&  0 
			}
	\right)
	\Psi_n (x - 3 a_0 /2 )
\right].
\end{eqnarray} 
\end{widetext}
In $k$-space this equation has the form 
$\varepsilon \Psi_n = H (n) \Psi_n$,
where the effective Hamiltonian for branch $n$ is:
\begin{eqnarray} 
H (n)
=
\left(
\matrix{
	0&   -t_{k_x n} \cr
	-t_{k_x n}^* & 0 \cr
	}
\right),
\\
t_{k_x n} 
= 
t \left[
		1 + 2 \cos \left(
				 \frac{ \pi n }{ M+2 }
			   \right)
		{\exp}	\left(
				- i \frac{3}{2}  k_x a_0
			\right)
  \right].
%%%%%%%%%%%%%%%%%%%%%%%%%%%%
\label{H_eff_k}%%%%%%%%%%%%%
%%%%%%%%%%%%%%%%%%%%%%%%%%%%
\end{eqnarray}
This Hamiltonian possesses accidental symmetry: the nanoribbon remains
unchanged under a shift by $3 a_0$ along the $x$-axis, yet the effective
Hamiltonian 
$H(n)$
is invariant under a shift by 
$3 a_0/2$. 
That is, the effective symmetry of the Hamiltonian is higher than the
geometric symmetry of the underlying system. This symmetry is destroyed when
the edge bonds are deformed. However, since in this paper we are interested in
the low-energy properties of the system, this peculiarity will play no role in
what follows.

Within our labelling scheme the same branch may appear under different values
of index $n$. Obviously, $n$ and $-n$ correspond to the same branch.
Furthermore, if 
$n>0$, $n'>0$, 
and
$n = - n' {\rm mod\ } (M+2)$,
both $n$ and $n'$ define the same branch. This means that there are $(M+1)$
independent branches:
\begin{eqnarray}
0 < n < M+2.
%%%%%%%%%%%%%%%%%%%%%%%%%%%
\label{band_index}%%%%%%%%%
%%%%%%%%%%%%%%%%%%%%%%%%%%%
\end{eqnarray} 
This is precisely the number of graphene unit cells in a nanoribbon unit
cell. The Hamiltonian of the nanoribbon is the direct sum of the
$H (n)$'s:
$
H = \sum_n H (n).
%%%%%%%%%%%%%%%%%%%%%%%%%%%%%
%\label{direct}%%%%%%%%%%%%%%%
%%%%%%%%%%%%%%%%%%%%%%%%%%%%%
$

The dispersion associated with a specific branch is given by Eq.~
(\ref{graphene_energy}), where $k_y$ is fixed by Eq.~
(\ref{quantization}). In other words, a branch samples the function 
$\varepsilon_{\bf k}$ along the line
$k_y = {\rm const.}$

Let us now find out under which circumstances the gapless branches appear.
Branch $n$ [$n$ satisfies Eq. (\ref{band_index})] is gapless if the complex
equation (a system of two real equations)
$t_{k_x, n}=0$
has a root satisfying Eq.(\ref{nanoribbon_bz}).

Solving this system of trigonometric equations one can prove that a zero
eigenvalue at 
$k_x = 0$
appears when
$\cos ( \pi n / (M+2)) = - 1/2$. The latter condition is fulfilled if
the argument of the cosine is 
$2 \pi / 3$.
This is possible provided that
$n=n_0$, where:
\begin{eqnarray}
n_0 = \frac{2 (M + 2)}{3},
%%%%%%%%%%%%%%%%%%%%%%%%%%%%%%%%%
\label{n0}%%%%%%%%%%%%%%%%%%%%%%%
%%%%%%%%%%%%%%%%%%%%%%%%%%%%%%%%%
\end{eqnarray}
and $n_0$ is an integer. In other words, the gapless branch is present only
when 
$(M + 2)$
is divisible by 3. This agrees with the conclusions of Ref. \cite{single_eIV},
where the single-electron calculations for a nanoribbon with no edge
deformation was performed.

The effective Hamiltonian for the gapless branch is:
\begin{eqnarray}
H_0 
=
H (n_0)
=
\left(
\matrix{
	0&   -t_{k_x } \cr
	-t_{k_x }^* & 0 \cr
	}
\right),
\\
t_{k_x }
=
t \left[
		1 - {\exp}	\left(
					-i \frac{3 k_x a_0}{2}
				\right)
  \right],
\\
\varepsilon_{k_x \pm} 
=
\pm 2 t
\left|
	\sin \left(
				\frac{3}{4} k_x a_0 
		   \right)
\right|.
\end{eqnarray} 
The gapless branch is characterized by 
$k_y = 4 \pi / (3 \sqrt{3} a_0)$. 
Since 
${\bf k} = (0, 4 \pi / (3 \sqrt{3} a_0))$ 
is the location of the cone ${\cal K}'$ [see Eq. (\ref{K'})], one can say
that the gapless branch is found in the nanoribbon spectrum only when the
quantization condition Eq. (\ref{quantization}) allows for existence of the
branch passing through the cone ${\cal K}'$.

\section{Spontaneous generation of the gap}
\label{instability}

In this section we show that the electronic branch, which appears to be
gapless according to the calculations reported above, is, in fact, unstable
toward the spontaneous opening of the gap. We prove that the edge bond
deformation is one possible instability leading to the gap generation.

\subsection{Modification of the Hamiltonian due to edge deformation}

To establish such an instability we need to calculate the ground state energy
of the nanoribbon with edge bonds deformed as shown on Fig. \ref{edge-H}. 

To achieve this aim we first determine how the edge deformation affects
the Hamiltonian of the gapless branch. We denote by
$\delta\! H = \delta\! H_{\rm l} + \delta\! H_{\rm h}$ 
the contribution to the Hamiltonian due to edge deformation. The subscript
`l' (`h') corresponds to a bond deformation at the lower (higher) edge of the
nanoribbon (see Fig. \ref{ribbon}).

The matrix element of 
$\delta\! H_{\rm l}$
between two states is equal to:
\begin{eqnarray}
\left< \Phi \right|
			\delta\! H_{\rm l}
\left| \Psi \right>
=
- \delta t \sum_m
			\phi^{{\cal A}*}_{{\bf R}_{m}}
			\psi^{\cal B}_{{\bf R}_{m}}
			+
			{\rm c.c.}
%%%%%%%%%%%%%%%%%%%%%%%%%%%%%%%
\label{matrix_element}
\end{eqnarray}
The summation in this formula runs over the deformed bonds at the lower edge:
${\bf R}_{m} = (3 a_0 m, 0)$.

Substituting in Eq. (\ref{matrix_element}) wave functions consistent with
Eq. (\ref{bc}), i.e.,
\begin{eqnarray}
\Psi_{\bf R} = \Psi 
{\rm e}^{- i k_x x}
\sin 
	\left[
		\frac{ 2 \pi n_0 }{ \sqrt{3} (M+2) a_0 }
		\left(
			y + \frac{\sqrt{3} a_0}{2}
		\right)
	\right]
\\
\nonumber
=
\Psi 
{\rm e}^{- i k_x x}
\sin 
	\left(
		\frac{4 \pi y}{3 \sqrt{3} a_0}
		+ \frac{2 \pi }{3}	
	\right),
\end{eqnarray}
we find:
\begin{eqnarray}
\left< \Phi \right|
			\delta\! H_{\rm l}
\left| \Psi \right>
&=&
\Phi^\dagger 
\left(
	\matrix{
		0&	-\delta t \cr
		-\delta t&	0 
		}
\right)
\Psi
\sin^2 (2\pi / 3)
%%%%%%%%%%%%%%%%%%%%%%%%%%%%%%%%%%%%%
\label{matrix_el1}%%%%%%%%%%%%%%%%%%%
%%%%%%%%%%%%%%%%%%%%%%%%%%%%%%%%%%%%%
\\
\nonumber
&\times&  \sum_m {\exp}	\left[
				-3 i  a_0 (k_x - k_x') m
			\right].
\end{eqnarray}
Therefore, for
$k_x^{\vphantom{'}}, k_x'$
satisfying Eq.(\ref{nanoribbon_bz}), Hamiltonian 
$\delta\! H_{\rm l}$
equals to:
\begin{eqnarray}
\delta\! H_{\rm l} 
=
-\frac{3 \delta t }{ 8 M + 8 }
\left(
	\matrix{
		0  &  1 \cr	
		1  &  0 
		}
\right)
\delta _{k_x^{\vphantom{'}}, k_x'},
%%%%%%%%%%%%%%%%%%%%%%%%%%%%%%
\label{h_l}%%%%%%%%%%%%%%%%%%%
%%%%%%%%%%%%%%%%%%%%%%%%%%%%%%
\end{eqnarray}
where
$2(M+1)$
in the denominator comes from the wave function normalization.

It is trivial to demonstrate that
$\delta\! H_{\rm l} = \delta\! H_{\rm h}$.
Superficially, this identity appears to be incorrect: clearly, there should
be a difference between the lower and higher edges at the level of the
Hamiltonian. However, we must remember that the expression Eq.~(\ref{h_l})
is not the total Hamiltonian 
$\delta\! H_{\rm l}$ 
(which is indeed different from 
$\delta\! H_{\rm h}$),
but rather its projection on the subspace spanned by a specific branch. These
projections cannot discriminate between the lower and the higher edge.

The total edge Hamiltonian 
$\delta\! H$
is equal to twice
$\delta\! H_{\rm l}$.
%\begin{eqnarray}
%\delta\! H = 2\delta\! H_{\rm l}.
%\end{eqnarray}
Thus, the Hamiltonian for a nanoribbon with the deformed edge bonds is:
\begin{eqnarray}
H_0 + \delta\! H
=
\left(
	\matrix{
			0&	-\delta t_{\rm eff} - t_{k_x} \cr
			-\delta t_{\rm eff} - t_{k_x}^* &	0
		}
\right),
\\
\delta t_{\rm eff}
=
\frac{3 \delta t }{ 4 M + 4 },
\end{eqnarray}
whose dispersion is given by
$
\varepsilon_{k_x\pm} 
=
\pm
\varepsilon (k_x)
$, where:
\begin{eqnarray} 
\varepsilon (k_x)
&=&
\sqrt{
	\left(
		\delta t_{\rm eff}
	\right)^2
	+
	4 t (t + \delta t_{\rm eff})
	\sin^2 ( 3 k_x a_0 / 4 )
},
\end{eqnarray}
which has a gap
$\Delta = 2 |\delta t_{\rm eff}|$.

Since our formerly gapless branch now acquired the gap, it might be confusing
to refer to such branch as `gapless'. Instead, we will now call it `$n_0$
branch', where $n_0$ is given by Eq. (\ref{n0}).

Heuristically, one can say that the gap appears because the boundary
conditions at the edges have changed. Indeed, we explained that the gapless
branch exists because the quantization condition Eq.~
(\ref{quantization}) makes this branch pass through the Dirac cone. When the
edge structure is altered, the boundary conditions are altered as a result.
The latter induces a modification of the quantization rule. Thus, in 
$k$-space, the $n_0$ branch shifts slightly off the cone pinnacle's location
and acquires a gap.

\subsection{Edge instability}
\label{instability_subsection}

To demonstrate the existence of an edge-induced instability we need to
calculate the energy of the $n_0$ branch:
\begin{eqnarray}
\epsilon_{\rm el}/L  =
- 2 \int_{-\pi/(3 a_0)}^{\pi/(3 a_0)} 
	\varepsilon (k_x )
	\frac{d k_x}{2\pi}.
\end{eqnarray} 
It is easy to show that:
\begin{eqnarray}
\epsilon_{\rm el}/L 
\approx
\epsilon^0_{\rm el}/L
-
\frac{2 (\delta t_{\rm eff})^2}{3 \pi t a_0}
\left(
	\ln \left| 
			\frac{t}{\delta t_{\rm eff}}
	    \right|
+
{\rm const.}
\right),
\end{eqnarray}
where $\epsilon^0_{\rm el}$ is the ground state energy calculated at
$\delta t_{\rm eff} = 0$.
The next term is the most singular correction to $\epsilon_{\rm el}^0$ due to 
$\delta t_{\rm eff}$.
This correction is not analytic in 
$\delta t_{\rm eff}$.

One also needs an expression for the nanoribbon lattice energy due to the
bond deformation:
\begin{eqnarray}
\epsilon_{\rm b}/L 
= 
2 \times \frac{1}{3 a_0} \times \frac{ \kappa u^2 }{ 2 } 
=
\frac{1}{3 a_0} \kappa u^2,
\end{eqnarray}
where $u$ is the variation in the edge bond length, $\kappa$ is the
stiffness of the bond (the energy 
$\epsilon_{\rm b} / L$
is composed of the energy of two deformed bonds per unit cell of the
nanoribbon; the energy of a single deformed bond is 
$\kappa u^2/2)$.

To proceed further we need to know how to relate the deformation of the
bond $u$ and 
$\delta t$.
Such information may be extracted from the quantum chemical calculations
\cite{carbon}.
However, to demonstrate that our system is unstable it is enough to assume
that at small $u$ we have
$\delta t \sim u$.
Then one can write the following expression for 
$\epsilon_{\rm b}$:
\begin{eqnarray}
\epsilon_{\rm b} / L
=
\zeta \frac{(M+1)^2}{ a_0 }
 \left( 
	\delta t_{\rm eff}
\right)^2,
\end{eqnarray}
where $\zeta$ is a phenomenological constant.

Finally, it is straightforward to check that the total nanoribbon energy 
$
(\epsilon_{\rm el}^0 + \epsilon_{\rm el}  + \epsilon_{\rm b})
$
has two minima at 
$\delta t_{\rm eff} = \pm \delta t^*$,
where:
\begin{eqnarray}
\delta t^*
\sim
t \exp \left[
		- 3 \pi t \zeta (M+1)^2
	\right] 
\ne 0.
%%%%%%%%%%%%%%%%%%%%%%%%%%%%%%%%%%
\label{t*}%%%%%%%%%%%%%%%%%%%%%%%%
%%%%%%%%%%%%%%%%%%%%%%%%%%%%%%%%%%
\end{eqnarray}
This expression shows that the nanoribbon energy is smallest when the bonds
at the edges are deformed and the $n_0$ branch has a gap.

Note that the calculations presented above rely on the mean-field
approximation. The latter is applicable to our one-dimensional system since
the order parameter (the bond deformation) is Ising-like. Thus, no Goldstone
mode is present, and we do not have to worry about critical fluctuations.

Above we demonstrated that an armchair nanoribbon of any width has a gap in
its electronic spectrum. There are two caveats to our discussion, however.

First, we proved that for our system at least one gap-opening instability
exists. We did not prove, yet, that the discussed mechanism is the only
possible path to generate the spectral gap. For example, the
electron-electron interaction can induce a gap
\cite{sandler}. 
Ultimately, the strongest instability must be determined by comparing the 
energies associated with particular mechanisms. The energy 
$\delta t^*$
characterizes the strength of the edge deformation instability. This energy
scale quickly vanishes for wider nanoribbons ($M$ is large) or ``stiff" edge
bonds ($\zeta$ is large). Under these conditions other mechanisms might be
important.

Second, due to special properties of the carbon atoms at the edge, it is
likely that the edge bonds are deformed by chemical forces, which are more
powerful than any intrinsic instability, including the one we have
discussed. First-principle numerical simulations support this point:
analysing Fig.~3 of Ref.~\cite{louieI}, we note that the edge bonds are
deformed even in nanoribbons where the unstable $n_0$ branch is absent;
moreover, the deformation magnitude is independent of the nanoribbon width.

One can say that our order parameter 
$\delta t_{\rm eff}$
is coupled to the fictitious ``field", which takes the system away from 
criticality (see Table \ref{parallel}).
%\begin{center}
\begin{table}
\begin{tabular}{||c|c|c||}
\hline
 		& Ising model 		& Nanoribbon 			\\ 
\hline\hline
Order 		& Magnetisation 	& Bond deformation $u$,		\\
parameter 	& $S$ 			&$u \sim \delta t$ 		\\
\hline
External field 	& Magnetic field $H$ 	& ``Edge field" $f$		\\
\hline
Energy 		& $a S^2 + b S^4 - HS$ 	& $a u^2 \ln u + b u^2 - fu $	\\
\hline\hline
\end{tabular}
\caption{
Analogy between the Ising model and the nanoribbon of graphene.
}
%%%%%%%%%%%%%%%%%%%%%%%%%%%%%%%%%%%%%%
\label{parallel}%%%%%%%%%%%%%%%%%%%%%%
%%%%%%%%%%%%%%%%%%%%%%%%%%%%%%%%%%%%%%
\end{table}
%\end{center} 
In such a situation
$\delta t_{\rm eff} \ne 0$; however, the value of
 $\delta t_{\rm eff}$
and corresponding spectral gap is determined not by Eq.~(\ref{t*}) but rather
by the strength of the external ``edge field".

Thus, the nanoribbon's spectral properties are controlled by the chemical
structure of the edges. If we find a way to vary the effects of the chemical
edge structure, we may tune the electronic properties of the nanoribbon
to our needs.
 
This problem can be dealt with the help of three different approaches. First,
one can try to hand-pick passivating radicals to guarantee preservation of
the edge structure. Second, one can treat different edges of the nanoribbon
with different radicals, which deform the edge bonds in opposite direction.
Clearly, these two proposals require considerable experimental work. A
third approach seems more promising: to close effectively the gap it is
enough to disorder the ``edge field". We will examine this idea
in the next section.

\section{Nanoribbon with edge disorder}
\label{disorder}

As we have seen in the previous section, edge bond deformations can open
a gap in the electronic spectrum of an armchair nanoribbon. The magnitude of
the gap is determined by the chemical properties of the passivating radical
at the nanoribbon's edges. In this section we investigate how the chemical
disorder at the edges affects the nanoribbon's spectrum. We will see that
the disorder effectively weakens the ``edge field". 

We consider a nanoribbon whose edges are treated by two different radicals,
`$\alpha$' and `$\beta$', which distort the bonds in opposite directions:
\begin{eqnarray}
\delta t_{\alpha} > 0,
\\
\delta t_{\beta} < 0,
\\
|\delta t_{\alpha} | = |\delta t_{\beta} | = \delta t_0.
%%%%%%%%%%%%%%%%%%%%%%%%%%%%%%%%%%%
\label{tA=tB}%%%%%%%%%%%%%%%%%%%%%%
%%%%%%%%%%%%%%%%%%%%%%%%%%%%%%%%%%%
\end{eqnarray}
When these radicals randomly attach to the nanoribbon's edges, the ``edge
field" becomes disordered. By adjusting the concentrations 
$n_{\alpha, \beta}$
of the two radicals, it is possible to vary the relative strengths of 
$\overline{\delta t}$
and  
$\delta t_{\rm dis}$:
\begin{eqnarray}
\overline{\delta t} 
= 
n_{\alpha} \delta t_{\alpha}
+
n_{\beta} \delta t_{\beta}
=
\delta t_0 (1 - 2 n_{\beta}),
%%%%%%%%%%%%%%%%%%%%%%%%%%%%%%%%%%
\label{dt}%%%%%%%%%%%%%%%%%%%%%%%%
%%%%%%%%%%%%%%%%%%%%%%%%%%%%%%%%%%
\\
\delta t_{\rm dis} (x) = \delta t (x) - \overline{ \delta t }.
\end{eqnarray} 
These two quantities are defined by Eq. (\ref{t_edge}).

The effective Hamiltonian for the $n_0$ branch in the presence of disorder is
equal to:
\begin{eqnarray}
H&=&H_0 + \delta\! H (x),
\\
\delta\! H (x)
&=&
-\, \frac{3}{ 4 M + 4 } 
\left(
	\matrix{
			0 & 1 \cr
			1 & 0 
		}
\right)
[\; \overline{\delta t} + \delta t_{\rm dis} (x)].
\end{eqnarray}
The coordinate representation of 
$\delta\! H (x)$ 
may be obtained through a procedure similar to the derivation of 
$\delta\! H_{\rm l}$ 
in Sect.~\ref{model}. Below we will assume a Gaussian distribution law for
the random quantity
$\delta t_{\rm dis}$ 
with the correlation function:
\begin{eqnarray}
\langle
	\delta t_{\rm dis} (x)
	\;
	\delta t_{\rm dis} (x')
\rangle
=
\sigma^2
f((x-x')/a),
\end{eqnarray}
or, in Fourier space:
\begin{eqnarray}
\langle
	\delta \hat t_{{\rm dis}, k_x}
	\;
	\delta \hat t_{{\rm dis}, -k_x'}
\rangle
=
\sigma^2 a L 
\;
\hat f(a k_x )
\;
\delta_{k_x, k_x'}.
\end{eqnarray} 
Here:
\begin{eqnarray}
\sigma^2
\sim
\left(
	\delta t_0
\right)^2
(n_{\beta} - n_{\beta}^2),
\end{eqnarray} 
sets the scale for the disorder strength variation, the scale $a$ is the
disorder correlation length. The function
$f(z)$
is a ``broadened $\delta$-function". It is even and non-negative, vanishes
quickly for
$|z|>1$.
In addition, this function is normalised by the condition:
\begin{eqnarray}
\int_{-\infty}^{+\infty} dz f(z) = 1.
\end{eqnarray}
Its Fourier transform $\hat f$ satisfies:
\begin{eqnarray}
\hat f (0) = 1.
\end{eqnarray}
In the plane-wave basis, the Hamiltonian 
$\delta\! H$ 
can be written as:
\begin{eqnarray}
\delta\! H
=
- \, \frac{3}{4M + 4}
\left(
	\matrix{
		0&	1 \cr
		1&	0 
		}
\right) 
\left(
	\overline{\delta t} \; \delta_{k_x^{\vphantom{'}}, k_x' }
	+
	\sqrt{\frac{\sigma^2 a}{L}} \, \tau_{k_x^{\vphantom{'}}, k_x' }
\right),
%%%%%%%%%%%%%%%%%%%%%%%%%%%%%%%%%%%
\label{deltaH}%%%%%%%%%%%%%%%%%%%%%
%%%%%%%%%%%%%%%%%%%%%%%%%%%%%%%%%%%
\\
\tau_{k_x^{\vphantom{'}}, k_x' }
=
\frac{1}{\sqrt{\sigma^2 a L}} 
\delta t_{{\rm dis}, k_x - k_x'},
%%%%%%%%%%%%%%%%%%%%%%%%%%%%%%%%%%
\label{tau}%%%%%%%%%%%%%%%%%%%%%%%
%%%%%%%%%%%%%%%%%%%%%%%%%%%%%%%%%%
\\
\langle
	|\tau_{k_x^{\vphantom{'}}, k_x' }|^2
\rangle
=
\hat f ( a (k_x^{\vphantom{'}}-k_x') ).
\end{eqnarray}
In Eqs.~(\ref{deltaH}) and (\ref{tau}), the length $L$ appears in the
denominator due to wave function normalization. In these equations we defined
the dimensionless random field 
$\tau_{k,k'}$
to show explicitly how the disorder matrix elements scale with the
nanoribbon length $L$ \cite{malyshev}.

As we see from Eq.~(\ref{deltaH}), the disordered and homogeneous parts of
the edge Hamiltonian enter as two different terms. Each term induces a
specific modification of the nanoribbon's spectrum: 
{\it the ordered part opens a gap, the disorder part localizes the wave
functions.}
The localization, however, is a weaker phenomenon than the gap generation.
Intuitively, this sounds quite reasonable: the effects of disorder may
``average out" to zero, while the ordered term acts ``coherently" over the
whole sample length. To make this statement rigorous we will apply
perturbation theory in orders of
$\delta\! H$. 
We will show that for a nanoribbon of a certain length the disorder may be
treated with the help of perturbation theory, while the ordered term
may not. To prove this we separately consider the two pieces of $\delta\! H$.

\subsection{Homogeneous edge deformation $\overline{\delta t}$}

Perturbation theory is applicable when the level spacing 
$\delta \epsilon \sim a_0 t / L$ 
is much larger than the matrix elements of 
$\delta\! H$.

To establish applicability range of the perturbation theory in orders of 
$\overline{\delta t}$,
we have to compare 
$\delta \epsilon$
with the gap
$\Delta$.
Thus, perturbation theory works if:
\begin{eqnarray}
L
\ll
l_{\rm gap} 
=
a_0
(M+1) \left(t/\overline{\delta t}\right).
%%%%%%%%%%%%%%%%%%%%%%%%%%%%%%%%%%
\label{pert_order}%%%%%%%%%%%%%%%%
%%%%%%%%%%%%%%%%%%%%%%%%%%%%%%%%%%
\end{eqnarray}
The length scale 
$l_{\rm gap}$
is analogous to the Compton length 
$l_{\rm C} = 1/(m_e c)$ 
of the Dirac equation.

When the disorder is absent 
($n_{\beta} = 0$), 
the gap has its biggest value, and 
$l_{\rm gap}$ 
is the shortest:
\begin{eqnarray}
l_{\rm gap}^{\rm min} = 
a_0 (M+1)
\left(t/\delta t_{0} \right).
\end{eqnarray}
Thus, one can say that perturbation theory in 
$\overline{\delta t}$
works for any value of
$n_{\beta}$
if
\begin{eqnarray}
L
\ll
l_{\rm gap}^{\rm min}.
\end{eqnarray}

\subsection{Disordered edge deformations $\delta t_{\rm dis}$}

The perturbation theory in 
$\delta t_{\rm dis}$
is applicable when:
\begin{eqnarray}
\frac{a_0 t}{L}
\gg
\frac{1}{M+1}
\sqrt{
	\frac{\sigma^2 a}{L}
     }.
\end{eqnarray}
This inequality may be transformed into:
\begin{eqnarray}
L
\ll
l_{\rm loc}
=
(M+1)^2
\frac{a_0^2}{a}
\frac{t^2}{\sigma^2}.
%%%%%%%%%%%%%%%%%%%%%%%%%%%%%%%%%%
\label{pert_disorder}%%%%%%%%%%%%%
%%%%%%%%%%%%%%%%%%%%%%%%%%%%%%%%%%
\end{eqnarray}
The scale 
$l_{\rm loc}$ 
is the localization length. (Our treatment of the disordered regime follows
closely Ref. \cite{malyshev}.)

When 
$\overline{\delta t} = 0$
(or, equivalently, 
$n_{\alpha} = n_{\beta} = 1/2$)
the disorder is the strongest. We can define the shortest possible
localization length 
$l_{\rm loc}^{\rm min}$.
It can be estimated as follows. If
$n_{\alpha} = n_{\beta} = 1/2$, 
$\sigma$
has its largest possible value:
\begin{eqnarray}
\sigma^{\rm max} = \delta t_0 ,
%%%%%%%%%%%%%%%%%%%%%%%%%%%%%%%
\label{sigma}%%%%%%%%%%%%%%%%%%
%%%%%%%%%%%%%%%%%%%%%%%%%%%%%%%
\\
l_{\rm loc}^{\rm min}
=
(M+1)^2
\frac{a_0^2}{a}
\frac{t^2}{(\delta t_0)^2}.
\end{eqnarray}
Therefore, if the sample length satisfies:
\begin{eqnarray}
L \ll l_{\rm loc}^{\rm min},
%%%%%%%%%%%%%%%%%%%%%%%%%%%%%%%%%%
\label{pert_disorder_all}%%%%%%%%%
%%%%%%%%%%%%%%%%%%%%%%%%%%%%%%%%%%
\end{eqnarray} 
the perturbation theory in the disorder strength is justified for any
concentration 
$n_{\beta}$.

\subsection{Conductance of a mesoscopic sample}

It is possible to prove that 
$l_{\rm gap}^{\rm min} \ll l_{\rm loc}^{\rm min}$. 
Indeed, this inequality is equivalent to:
\begin{eqnarray}
(M+1)
\frac{t }{\delta t_0} 
\gg
\frac{a}{a_0}.
\end{eqnarray}
Both factors on the left-hand side of this expression are much larger
than unity. Therefore, unless $a$ is very big, 
$l_{\rm gap}^{\rm min} \ll l_{\rm loc}^{\rm min}$. 
Loosely speaking, this inequality suggests that the disordered field is a
much weaker perturbation than the ordered one.

Consider now a nanoribbon whose length $L$ satisfies:
\begin{eqnarray}
l_{\rm gap}^{\rm min} \ll L \ll l_{\rm loc}^{\rm min}.
%%%%%%%%%%%%%%%%%%%%%%%%%%%%%%%%%%%%%%
\label{meso}%%%%%%%%%%%%%%%%%%%%%%%%%%
%%%%%%%%%%%%%%%%%%%%%%%%%%%%%%%%%%%%%%
\end{eqnarray}
This means that even the strongest disorder
($l_{\rm loc} = l_{\rm loc}^{\rm min}$)
cannot create a well-developed localization in our nanoribbon; on the other
hand, when the sample is close to perfect order 
($n_{\beta} \ll 1/2$)
the spectral gap fully manifests itself. Let us now study the electrical
conductance of such nanoribbon.

Assume first that our system has no edge disorder: 
$\delta t(x) = \delta t_{\alpha}$.
Then 
$l_{\rm gap} = l_{\rm gap}^{\rm min}$, 
and Eq. (\ref{pert_order}) is violated. Therefore, the ``edge field" opens a
gap in the spectrum. The dimensionless conductance $g$ of a sample with the
gap is exponentially small at $T=0$:
\begin{eqnarray}
\ln g \sim - \, \frac{L}{l_{\rm gap} }
= 
- \, \frac{1}{M+1} \frac{\overline{\delta t}}{t} \frac{L}{a_0}.
%%%%%%%%%%%%%%%%%%%%%%%%%%%%%%%%%%%%%
\label{g_order}%%%%%%%%%%%%%%%%%%%%%%
%%%%%%%%%%%%%%%%%%%%%%%%%%%%%%%%%%%%%
\end{eqnarray}
When we slightly disorder our system by introducing a small concentration 
$n_{\beta}$ 
of radicals `$\beta$', the conductance increases since
$\overline{\delta t}$ 
decreases [see Eq.(\ref{dt})]. Thus, 
{\it the disorder improves electrical conductance!}

In the opposite limit of complete disorder, we have
$\overline{\delta t} = 0$,
and the localization length becomes
$l_{\rm loc} = l_{\rm loc}^{\rm min}$.
Perturbation theory may now be applied since Eq.~(\ref{pert_disorder})
holds true. Instead of a disorder-induced localization, which is a
non-perturbative phenomenon, in a sufficiently short nanoribbon the disorder
creates weak corrections to the properties of $H_0$. In such sample the
conductance remains finite even at $T=0$.

Thus, we reach the following counter-intuitive conclusion:
{\it a completely ordered nanoribbon shows ``insulating" behaviour, while a
disordered one shows ``metallic".}
[We put quotes around ``insulating" and ``metallic" for metal and insulator
are quantum phases, which can be unambiguously defined only in the
thermodynamic limit 
$L \rightarrow \infty$;
however, the latter limit is incompatible with Eq. (\ref{meso}).] 

As the system moves from perfect order to total disorder, it passes through
a crossover from ``insulating" to ``metallic" conductance. Indeed,
Eq.~(\ref{g_order}) is applicable only when 
$L/l_{\rm gap}$
is much bigger than unity. If
\begin{eqnarray}
L/l_{\rm gap} \sim 1 \Leftrightarrow g \sim 1,
\end{eqnarray}
one can validate the perturbation theory not only in orders of
$\delta t_{\rm dis}$,
but also in orders of
$\overline{\delta t}$
[see Eq.(\ref{pert_order})]. Therefore, once $L$ exceeds 
$l_{\rm gap}$,
the exponential dependence of $g$ is replaced by a slower function, and
$g$ remains of order unity down to the completely disordered regime.

\subsection{Conductance of a long nanoribbon}

Finally, let us comment on the conductance behaviour in a thermodynamically
large sample, whose length satisfies:
\begin{eqnarray}
l_{\rm gap}^{\rm min} < l_{\rm loc}^{\rm min} \ll L.
%%%%%%%%%%%%%%%%%%%%%%%%%%%%%%%%%%%%
\label{hier}%%%%%%%%%%%%%%%%%%%%%%%%
%%%%%%%%%%%%%%%%%%%%%%%%%%%%%%%%%%%%
\end{eqnarray}
In this situation both terms of 
$\delta\! H$ 
cannot always be treated with the help of perturbation theory. Consequently,
the conductance is always exponentially suppressed: when there is perfect
order, the conductance follows Eq.(\ref{g_order}); in the opposite case
(complete disorder) we have:
\begin{eqnarray}
\ln g \sim - \, \frac{L}{l_{\rm loc}}.
%%%%%%%%%%%%%%%%%%%%%%%%%%%%%%%%%%%%
\label{g_dis}%%%%%%%%%%%%%%%%%%%%%%%
%%%%%%%%%%%%%%%%%%%%%%%%%%%%%%%%%%%%
\end{eqnarray}
The latter equation is a manifestation of wave function localization,
which can only be observed in a sample whose length exceeds 
$l_{\rm loc}$.

Comparing Eqs.~(\ref{g_dis}) and (\ref{g_order}) with the help of Eq.~
(\ref{hier}), we note that, as well as in the case of a short nanoribbon,
the conductance of a perfectly ordered sample is much smaller than the
conductance of a totally disordered sample.

\begin{figure} 
\centering
\leavevmode
\epsfxsize=8.5cm
\epsfysize=7.5cm
\epsfbox{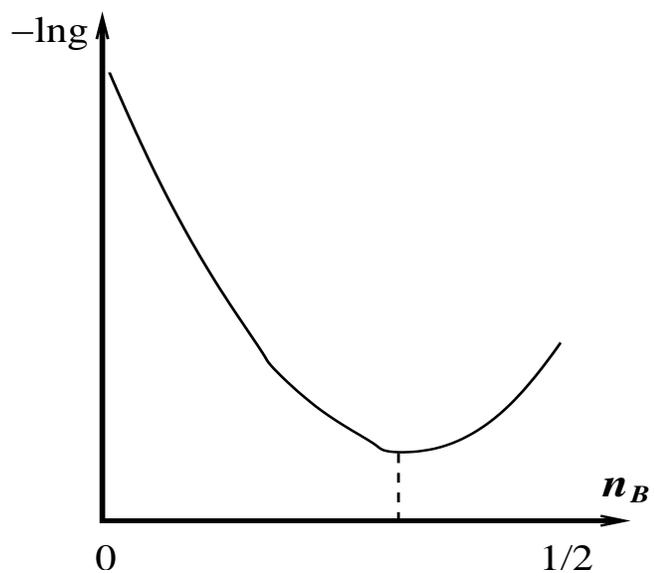}
\caption[]
{\label{conduct}
Qualitative behaviour of the conductance of the long nanoribbon as a
function of disorder ($n_{\beta}$ is the concentration of the disordering
radical).  The dashed line marks the crossover from the gap-dominated to
the disorder-dominated regime. Note the counter-intuitive trend left of the
dashed line: the conductance is {\it increasing} with increasing disorder!
}
\end{figure}

The conductance $g$ of a long nanoribbon is a non-monotonous function of
disorder (see Fig.~\ref{conduct}). Such a behaviour is a consequence of
the crossover from the gap-dominated to the disorder-dominated regime. When
the concentration 
$n_{\beta}$ 
is small, the gap is the dominant parameter controlling the conductance.
Under such circumstances Eq.(\ref{g_order}) is obeyed. The disorder acts
mainly to reduce the gap.  Thus, if disorder is weak, then $g$ is an
increasing function of 
$n_{\beta}$. 

As 
$n_{\beta}$
keeps growing,
$l_{\rm gap}$ 
increases, while
$l_{\rm loc}$
decreases. The crossover occurs at:
\begin{eqnarray}
l_{\rm loc} \sim l_{\rm gap},
\end{eqnarray}
and the sample enters the disorder-dominated regime. The conductance is
given by Eq. (\ref{g_dis}). It is a decreasing function of 
$n_{\beta}$
when the latter is close to 1/2.

\section{Conclusions}
\label{conclusions}

In this paper we studied the effects of edge bond deformations on the
electronic properties of nanoribbons. We have seen that a nanoribbon of a
certain width is unstable with respect to a spontaneous deformation of the
edge bonds. While such deformation increases the energy of the affected
bonds, it also reduces the electronic energy. As a result of a `bond
instability' the electronic spectrum acquires a gap at the Fermi level.

We also pointed out that this instability is difficult to observe in a real
system. The culprit is the chemical structure of the edges, which deforms
the bonds to optimize the chemical energy at the edges. Although it might
be hard to remove this chemical modification of the edge bonds, it is
quite possible to reduce its effect on the electronic spectrum by
disordering the radicals passivating the edge bonds. We demonstrated that the
disorder can vary the electrical conductance of a nanoribbon. In case of
a short nanoribbon, the conductance would change from ``insulating" regime
at low disorder to a ``metallic" regime at high disorder. When the
nanoribbon's length is large, the conductance is a non-monotonous function of
the disorder. Thus, the disorder may be a useful tool which allows one to
control the electric transport through nanoribbons.

\section{Acknowledgements}

We gratefully acknowledge partial support from the National
Security Agency (NSA), Laboratory Physical Science (LPS), Army
Research Office (ARO), National Science Foundation (NSF) grant No.
EIA-0130383, JSPS-RFBR 06-02-91200, and Core-to-Core (CTC) program
supported by Japan Society for Promotion of Science (JSPS). SS
acknowledges support from the EPSRC via No. EP/D072581/1, 
EP/F005482/1, and ESF network-programme  ``Arrays of Quantum Dots and Josephson
Junctions''. AVR is grateful for the support provided by the Dynasty
Foundation and by the RFBR grant No. 06-02-16691.

%%%%%%%%%%%%%%%%%%%%%%----BIBLIOGRAPHY----%%%%%%%%%%%%%%%%%%%%%%%%%%%%%%%%%

\newpage

\end{document}